# Construction and Evaluation of Mandarin Multimodal Emotional Speech Database

Ting Zhu[1], Liangqi Li[1], Shufei Duan[1], Xueying Zhang[1], Zhongzhe Xiao[2], Hairong Jia[1], Huizhi Liang[3]

(1 *College of Information and Computer, Taiyuan University of Technology, Taiyuan, Shanxi* 030024, *China*;

2 *College of Optoelectronic Information Science and Engineering, Suzhou University, Suzhou, Jiangsu* 215006,*China*;

3 *School of Computing, Newcastle University, Newcastle NE17RU,United Kingdom*）

**Abstract:** A multi-modal emotional speech Mandarin database including articulatory kinematics, acoustics, glottal and facial micro-expressions is designed and established, which is described in detail from the aspects of corpus design, subject selection, recording details and data processing. Where signals are labeled with discrete emotion labels (neutral, happy, pleasant, indifferent, angry, sad, grief) and dimensional emotion labels (pleasure, arousal, dominance). In this paper, the validity of dimension annotation is verified by statistical analysis of dimension annotation data. The SCL-90 scale data of annotators are verified and combined with PAD annotation data for analysis, so as to explore the internal relationship between the outlier phenomenon in annotation and the psychological state of annotators. In order to verify the speech quality and emotion discrimination of the database, this paper uses 3 basic models of SVM, CNN and DNN to calculate the recognition rate of these seven emotions. The results show that the average recognition rate of seven emotions is about 82% when using acoustic data alone. When using glottal data alone, the average recognition rate is about 72%. Using kinematics data alone, the average recognition rate also reaches 55.7%. Therefore, the database is of high quality and can be used as an important source for speech analysis research, especially for the task of multimodal emotional speech analysis.

## Introduction

Speech is one of the most natural and effective ways for human beings to communicate their emotions, thoughts and other information [1], among which speech emotion recognition has received widespread attention and plays an indispensable role in human-computer interaction intelligent technology. For example, the driver's emotional state can be judged through the driver's voice, facial expression and physical behavior, and the driver can be reminded to stay awake in real time, so as to avoid traffic accidents [2]. In human-machine dialogue, mastering human emotional states can help intelligent robots make appropriate responses, express comfort and understanding in a timely manner, and improve user experience. In terms of speech translation, the combination of the speaker's voice emotion information and automatic translation results can enable speakers from all sides to communicate more smoothly [3]. The premise of speech emotion recognition research is the support of emotion speech database. The quality of an emotion speech database greatly affects the performance of the emotion recognition system trained by it. Due to factors such as the involvement of emotion in psychology, the differences in individual cognition and expression of emotion, the non-uniformity of emotional stimulation methods and experimental recording standards, most research institutions have established corresponding emotional speech databases according to their own scientific research needs, and there is no unified standard for establishing emotional speech databases [4].

The establishment of an emotional speech database with natural emotion and smooth expression has gradually become a research hotspot at home and abroad. There are many categories of emotion speech bank, which can be divided into performance type, guiding type and natural type according to the categories that stimulate emotion [5]. The expression of emotion can be divided into obvious emotional clues and implicit emotional stimuli [6]. According to the emotion description model, it can be divided into discrete emotion database and dimension emotion database. The discrete emotion database expressed emotion using adjective labels (happy, angry, afraid, etc.), while the dimensional emotion database expressed pleasure-arousal (VA) [7] or in coordinate form on the three-dimensional pleasure-arousal-dominance (PAD) [8], which is more capable and accurate in emotional representation. At present, most of the research fields at home and abroad are discrete emotion corpus. for example, the Institute of Automation, Chinese Academy of Sciences (CASIA), the Emotional Database for German Speech, EMO-DB and the guided Ryerson Audio-Visual Database of Emotional Speech and Song (RAVDESS), etc., while the domestic dimension data is relatively small. Nowadays, databases tend to be built to contain both discrete and dimensional annotation information. Therefore, it is necessary to build a high-quality affective speech database with real emotional stimulation, natural expression, and coexistence of discrete and dimensional annotation information.

The further integration of signal processing and artificial intelligence has facilitated the development of advanced intelligent systems that can detect and process emotional information contained in multimodal sources. However, at present, the most advanced model frameworks are all based on single mode generation, that is, text, audio or video, and these systems have limitations in meeting the requirements of robustness, accuracy and overall performance [9]. Relying only on single mode emotion recognition of speech can no longer meet the development needs of the overall trend. The purpose of multi-modal data fusion is to improve the accuracy and reliability of estimation, make up for the defect of insufficient information of a single mode, and have the advantages of mutual supervision and multi-dimensional perception in the human-computer interaction process. In this regard, a number of research groups have built multimodal databases, For example, Vera am Mittag German Audio-Visual Emotional speech database (VAM) [10] recorded a dual-modal emotional database including video and Audio through German TV talk shows; Emotion in Movies (EMOVIE) database containing text and speech was established for text-to-speech conversion [11]. At home, 3D Electromagnetic Articulography (3D electromagnetic articu-lography), Duke Kunshan University Jinan Uni-versity Electromagnetic Articulography, EMA (Duke Kunshan University Jinan Uni-Versity Electromagnetic articulography, DKU-JNU-EMA database [12], which aims to study the recognition of Mandarin and Chinese dialects; Ren Guofeng et al. [13] designed and established a Mandarin bimodal emotion database including EMA and audio. However, as far as we know, there is still no multi-modal database that combines EMA and voice simultaneously and includes video, Electroglottography (EGG) and dimensional annotation at home and abroad.

This paper designed and established a multimodal database containing seven emotions, namely neutral, happy, pleasant, angry, indifferent, sad and grief, and involving articulatory motor parameters, speech, glottic and facial micro-expressions. At the same time, the speech annotation of PAD was carried out. So, the discrete emotion database and dimensional emotion database were combined. It provides an important data base for the study of discrete emotion and dimensional emotion and the study of the relationship between vocal and motor organs.

# 1 Database design

The multi-modal audio-video emotional pronunciation database (Suzhou and Taiyuan emotional dataset on mandarin with electro-magnetic articulation, electrography, video and audio, STEM-E$^2$VA) is mainly composed of three parts, namely, the database of pronunciation kinematics and acoustics recorded by EMA-AG501, the glottis database recorded by EGG-D100 and the video database recorded by SONY-FDR, including voice signals, laryngeal contact signals, vocal organ (tongue and lips) motion trajectory parameters and facial video data of subjects.

In the current research of emotional phonetics, Ververidis et al. [14] summarized 32 existing emotional phonetics databases, which involve more than 10 languages and have different emotional categories and quantities. Most of them agree with the basic emotions: fear, happiness, anger, disgust, sadness and surprise. Therefore, this paper chooses neutral emotion and the three emotional states with the highest number of uses, namely Anger, Sadness and Happiness, and follows them. Chen Ying et al. [15] divided these three emotions into two categories: strength and weakness, and analyzed the relationship between different but similar emotions in the continuous dimensional space. Table 1 shows seven emotional types used in this database.

Table. 1  Type of emotion

| Neutral | Happy | | Angry | | Sad | |
|---|---|---|---|---|---|---|
| （No emotion） | strong | weak | strong | weak | strong | weak |
| | happy | pleasant | angry | indifferent | grief | sad |

## 1.1 Text design

In the performance database, when the voice is recorded, the participant uses the emotion to express the specified text content according to the given emotion label. However, most performative databases allow participants to perform emotionally without a stimulus source, which leads to a lack of authenticity in the collected data [16]. Emotional stimulation is carried out with the appropriate motivation source, and the appropriate performance in this emotional atmosphere will obtain real and natural data. This database combines the autobiographical recall emotion induction method [17] and the imaginative emotion induction method [18] to perform to a certain extent, and constructs and guides the specific contextual environment for emotional brewing according to the hypothetical content of the text. By bringing the participants into a constructed emotional environment or evoking the same emotional experiences as them, we obtained recordings of the participants' emotions that were close to the real situation.

Table. 2  Database corpus

| Type | Text |
|---|---|
| vowel | 1）Single vowel：a、o、e、i、u、ü; <br> 2）Compound vowels：ai、ei、ui、ao、ou、iu、ie、üe、er; <br> 3）Nasal rhyme：an、en、in、un、ün、ang、eng、ing、ong |
| statement | 1) wǒ kàn jiàn zhuō zi shàng yǒu yī gè fěn sè de xī hóng shì. <br> 2) mì fēng fēi qǐlái huìyǒu wēng wēng de sheng yīn. <br> 3) tā měi gè zhōu mò dōu huìnà er qù yī tàng. <br> 4) zài guò jǐ tiān jiù yào kāi shǐ luò yè le. <br> 5) tā men zhù de shì liùgè rén yī jiān wū zi de sù shè. <br> 6) qù nián qīyuè, tā hái qīn zìxiě le yī shǒu gē. <br> 7) kě shì zhè ge xué xiào yǒu jǐbǎi wèi lǎo shī. <br> 8) jīn nián de méi yǔ jì shí zài shì tǐng cháng de. <br> 9) tā hěn xǐ huan yòng gāng qín dàn huān lè sòng. <br> 10) xiàn zài yǐjīng hěn shǎo yǒu rén yòng cídài lù yīn jī le. <br> 11) qián tiān shítáng gēn běn jiù méi yǒu mài bāo zi. <br> 12) xiǎo lán jīng cháng pàn zhe kuài diǎn xià xuě. <br> 13) tā zuì xǐ huan chàng de gē shì zhù nǐ sheng rì kuài lè. <br> 14) nǐ měi cì qù yóu yǒng dōu bù ài tú fang shài shuāng. <br> 15) duì le, qián tiān dǎ léi de shí hou, mén qián yǒu yī zhī hǎi ōu. <br> 16) yán sè yīng gāi tú de gèng jūn yún diǎn jiù hǎo le. |

The text material of the database consists of finals and sentences, and the text material of the finals is shown in Table 2. Sentence selection is one of the key parts of the affective speech database, and excellent sentences can enable participants to better express their emotions, which is the first step to ensure the smooth progress of speech emotion recognition [19]. The selection of sentences in this database is considered from three aspects:

sentence type, sentence content and sentence length, and declarative sentences are used as corpus for sentence types, and modal words such as "?", "ah" and "ah" are avoided. The content of the sentence adheres to the principle of "neutral", does not use emotional words, emphatic words and adverbs of degree, etc., and selects sentences commonly used in life, so as to be closer to the real nature; It is stipulated that the sentence duration is about 5s under the normal speech rate. Table 2 shows the specific text content.

### 1.2 Selection of participants

Considering that the school's acting students express emotions in a rather exaggerated way, we believe that a well-trained actor is not the best choice for expressing natural emotions. Therefore, the database was recorded by recruiting students from non-acting departments of the school, allowing the participants to express real emotions as naturally as possible, thus ensuring the popularity and authenticity of the data. In the end, 22 non-performance college students (1:1 ratio of men and women) were selected to participate in the recording, and these students were composed of undergraduate and graduate students, aged between 21~27 years old. Prior to data collection, all participants were asked to undergo the Symptom Self-Rating Scale (SCL-90) survey and emotional expression training, and qualified subjects were informed of the procedures required for data collection and signed an informed consent form [20]. The selection of participants should follow the following principles:

1) Participants should be native Chinese speakers, standard Mandarin, no serious local accents, and be able to express themselves fluently.

2) The subject has no throat disease, and does not suffer from colds and other diseases that affect pronunciation on the day of recording, and the mood is normal, which will not affect the expression of emotions.

3) The participant's oral cavity has not been operated on, and the teeth should not be inlaid with metal products.

### 1.3 Experiment

#### 1.3.1 Acoustic-articulatory database

EMA-AG501[21] is an instrument specially designed for collecting kinematic data of vocal organs, which is introduced by Carstens Company in Germany. It includes 24 channels for collecting kinematic data with a sampling rate of 250 Hz and 1 channel for collecting audio data with a sampling rate of 45 kHz. This instrument collects the kinematic data of the vocal organs, using 13 sensors in total, among which 4 lip sensors (located at the upper and lower lip and the left and right corners of the mouth, denoted as L1-L4) and 3 tongue sensors (located at the tip of the tongue, the body of the tongue and the root of the tongue, denoted as T1-T3) collect the kinematic data. Because the head of the subject will inevitably move in the experimental stage, which will lead to the inaccuracy of the three-dimensional coordinate values in the Cartesian coordinate system under the magnetic field, the head is calibrated by three reference plane sensors (located on the left and right mastoid and nose bridge, denoted as B1-B3) and three bite plate sensors (denoted as P1-P3), as shown in Figure 1. During calibration, the subjects sit on the wooden chair under the AG501 instrument and bite the occlusal plate. A coordinate rotation matrix is formed by the motion data of the reference plane sensor and the occlusal plate sensor, and the space state is generated by the original Euclidean data on the AG501 controller. Simultaneously, the subjects are required to rotate their heads to form standard coordinates with the center of the occlusal plate as the coordinate origin, P1 and P2 on the Y axis and symmetrical with the origin, P3 on the X axis, and the Z coordinates of the three reference points are all zero.

After the head calibration data collection is completed, the remaining sensors are attached to the tip of the tongue, the middle of the tongue, the base of the tongue, the upper and lower lips and the left and right corners of the mouth respectively. The specific attachment positions of each sensor are shown in Figure 2. In order to get more accurate pronunciation data, the tip sensor is stuck at a distance of 1 cm from the physiological tip of the tongue, and the sensor at the base of the tongue is stuck to the base of the tongue as much as possible within the range that the speaker can bear, while the sensor at the middle of the tongue is located at the midpoint of the tip of the tongue and the sensor at the base of the tongue, and its distance from the two sensors should not be less than 1 cm.

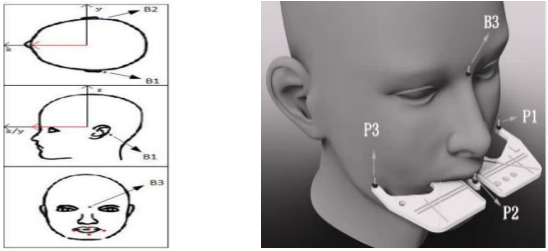

(a) Reference plane sensor settings    (b) Occlusal plate sensor setting

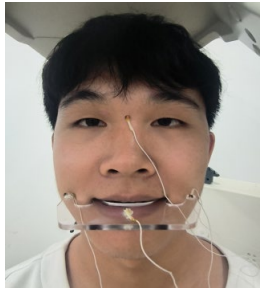

(c) Subject's head calibration

Fig.1 Calibrate the head sensor settings

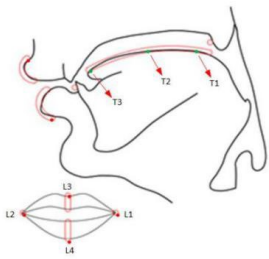 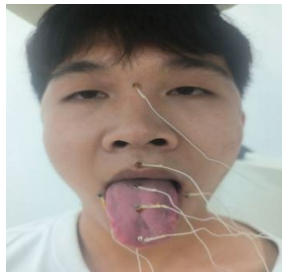

(a) Setting of the sensor of the pronunciation organ    (b) Schematic diagram of subject sticking

Fig.2　Collect data sensor settings

#### 1.3.2　Video database

The video data was recorded using the SONY FDR-AX100 in MP4 format of 3840x2160 pixels with a frame rate of 25 fps, a data rate of 54 Mbps and a total bit rate of 56 Mbps. The captured audio signal is two-channel with a bit rate of 1.5 Mbps. For the extracted WAV file, the signal is downsampled to 48 kHz (16 bits). Screenshots of the participants' facial micro-expressions from the video are shown in Figure 3.

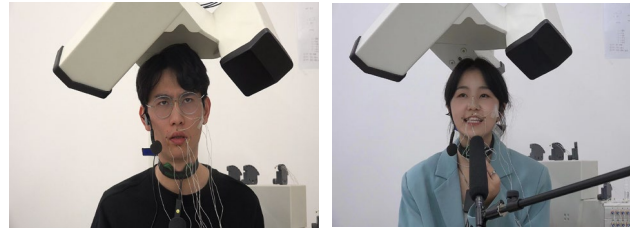

(a) An angry expression    (b) Happy micro-expression

Fig.3 Video micro-expression capture

#### 1.3.3　Glottic database

Electronic glottography is a non-invasive measurement device that collects voice signals [22] and has been widely used in the field of speech and pathological voice research. The electronic glottometer indirectly reflects the movement state of the vocal cords by measuring the current passing through the vocal cord area, which is simple to operate and harmless to the human body. Before the experiment, the recording staff wore the device to ensure that the electrodes were close to the glottis and did not affect the normal pronunciation of the participants, as shown in Figure 4.

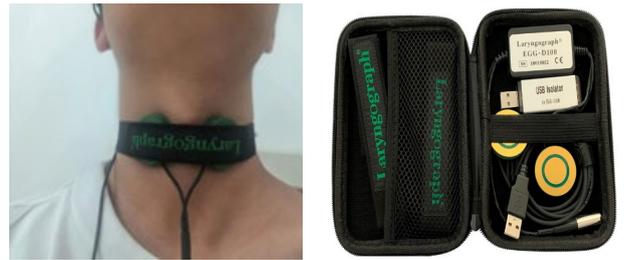

Fig.4 EGG wearing schematic and physical drawings

### *1.4　Data processing and screening*

During the recording, the participants need to express each vowel in the order of the four tones of yinping, yangping, upper voice, and devoice, and after the finals are recorded, the 16 sentences and texts will express 7 kinds of emotions in turn. In the process of data processing, we did not process the video data, considering that the EGG microphone and face sensor in the video data will cause a certain occlusion of the subject's face, and the required micro-expression pictures need to be manually captured. Users can edit the required data from the video according to their own needs.

1.4.1 Kinematic data preprocessing

The preprocessing steps for the articulatory motion data acquired by the EMA-AG501 are shown in Figure 5. Firstly, the original voltage amplitude data (.amp) is transformed into the original three-dimensional spatial data (.rawpos). Then, the reference plane sensor is used to generate three-dimensional spatial data. then convert it to ASCII code; Finally, with the help of Visartico, the trajectory of the articulation organs was observed, and if the complete lip shape and tongue order appeared, it was normal data, so as to filter out stable articulation data.

Root Mean Square Error (RMSE) is an important indicator used to measure the degree of data dispersion. Therefore, the three-dimensional spatial motion parameters of the upper and lower lips, the left and right lips and the tongue were also extracted, and the root mean square error was used as another evaluation index of the pronunciation kinematic data, the smaller the root mean square error value was the more stable the pronunciation motion data, and the speech with a value of ≤ 10 mm was judged to be stable data, and vice versa was filtered.

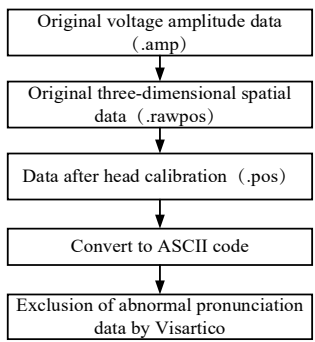

Fig.5　Pre-processing flow chart

1.4.2 Acoustic data processing

In order to obtain objective emotion data, we need to evaluate the effectiveness of the speech database to screen the voice quality and emotional expression. Therefore, the subjective discrimination method was used to score the data [23-25], and the evaluation indicators are shown in Table 3.

Speech intelligibility is the evaluation criterion of the recorded speech clarity, and it is not necessary to distinguish whether the emotional expression is sufficient and discernible, but the evaluation of emotional expression needs to inform the volunteers of the emotions that the participants in the current speech originally wanted to express. All data were scored separately according to speech affectivity and emotional expression, and neutral affectivity did not need to be evaluated for emotional expression. According to the results of data statistics, the threshold was set to speech intelligibility evaluation greater than 2.5 and emotional expression evaluation greater than 2.5, and the data that met the requirements were selected for subsequent analysis.

Table. 3　Acoustic data evaluation scale

| Scores | Grade | |
|---|---|---|
| | Speech intelligibility | Emotional expression |
| 3 | Noiseless and semantically clear | Strong emotional expression |
| 2 | Slightly noisy but semantically clear | Moderate emotional expression |
| 1 | Noticeable noise and interfere with semantic communication | Slight expressions of emotion |
| 0 | The audio is noisy and unintelligible | No emotion |

1.4.3 Glottic data processing

The signals collected by EGG are divided into left channel (glottis) and right channel (acoustics), and the left channel signal is positively correlated with the contact area of vocal cords [26]: when the contact area of vocal cords increases, the signal is enhanced; As the contact area decreases, the signal decreases. At the same time, the difficulty of obtaining glottic signals varies with the body structure of the subjects. Therefore, when screening glottis data, the data whose left channel signal is not obvious, unstable or disappeared are excluded.

1.4.4 Filter the results

After filtering the data of each modality, the results obtained are in Table 4. We have unified the numbering of this database, and the naming format of the data is "modality-participant-corpus type-corpus content". Among them, modalities are divided into A (acoustic), K (articulation), G (glottis); Participants were divided into M1-M11 (male) and F1-F11 (female). The corpus type is divided into f (finals) and s (sentence), and when the corpus type is sentence, the emotion type is divided into 0 (neutral), 1 (happy), 2 (pleasant), 3 (angry), 4 (indifferent), 5 (grief) and 6 (sad), and the corpus content is 01-16 (16 text sentences). If the corpus type is finals, there is no emotion type, and its corpus content is 01-24 (24 finals pinyin). For example, the number AM5s115 represents the acoustic data of the 5th male recording the 15th sentence under the happy emotion.

Table. 4  STEM-E$^2$VA database content

| Data type | Corpus type | Number of corpora | | | | | | | total |
|---|---|---|---|---|---|---|---|---|---|
| Pronunciation data | vowel | Single vowel | | Compound vowels | | Nasal rhyme | | | 528 |
| | | 132 | | 198 | | 198 | | | |
| Acoustic data | sentence | neutral | happy | Pleasant | angry | indifferent | grief | sorrowful | 2415 |
| Laryngeal data | | 352 | 335 | 345 | 345 | 352 | 337 | 349 | |
| Video data | | | | | | | | | 24.5h |

## 2 PAD annotation and analysis

### 2.1  Scale model

The three-dimensional affective model of PAD was proposed by Russell et al. [8] in 1974 in combination with previous psychological research, and they divided emotion into three dimensions: pleasure, arousal, and dominance, as shown in Figure 6. Pleasure-displeasure (denoted by P) indicates the positive and negative characteristics of emotions, that is, the negative and positive states of emotions. Arousal (arousal-nonarousal, denoted by A) indicates the level of emotional physiological arousal and alertness, i.e., whether an individual's response to the environment is excited or depressed. Dominance-submissiveness is represented by D, which indicates the control and influence of emotions on others and the external environment, and distinguishes whether the emotional state is generated by the individual's subjective or objective environment.

In the PAD affective model, discrete emotions can be represented by a series of coordinates, which are labeled by the PAD affective scale revised by the Chinese Academy of Sciences. Each item of the Nine-Point Semantic Difference Scale [27] consists of two pairs of opposite adjectives spaced 9 segments apart, and each dimension contains 4 items. In order to more accurately reflect the emotions, Gong et al. [28] supplemented the 7-point semantic difference scale on the basis of the improved 5-point difference scale. In order to make it easier for participants to understand and score, the original 7-point value -3 to 3 was changed to 1 to 7, and the 4 items in 3 dimensions were grouped together. For example, if the participant hears one of the emotions of anger, contempt, grief, or anger during the labeling, the pleasure of the voice is rated as 1, as shown in Table 5.

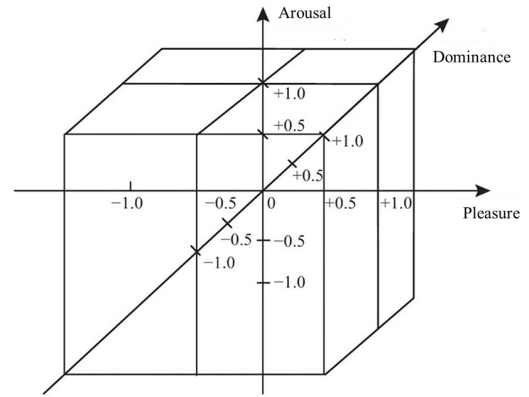

Fig.6   PAD schematic diagram

Table. 5 Simplified PAD mood scale

| Meaning | Emotional type | | | | | | |
|---|---|---|---|---|---|---|---|
| | 1 | 2 | 3 | 4 | 5 | 6 | 7 |
| Pleasure | Angry | Angry | Unpleasant | Neutral | attentive | Concern | Interested |
| | pejorative | Cold | Calm | Neutral | Mild | Friendly | Intimate |
| | Grief | Sad | perturbed | Neutral | Satisfied | Satisfied | Delighted |
| | Infuriated | sorrow | Boredom | neutral | Pleasant | Happy | Excited |
| Arousal | Sleepy | Lazy | Absent-minded | Neutral | advertent | attentive | Sober |
| | Calm | Relaxed | Mild | Neutral | Interested | Cheerful | Excited |
| | Depressed | Relaxing | Peaceful | Neutral | Comfortable | Attention-grabbing | Interested |
| | Restrained | Scared | Worried | neutral | Stunned | Surprised | Surprised |
| Dominance | Controlled | Anxious | perturbed | Neutral | Satisfied | Exhilarating | Mastered |
| | compliant | Accepted | Agreed | Neutral | Hopefully | Covetous | Dominative |
| | Humility | Shy | dependent | Neutral | Confident | Conceited | Proud |
| | Affected | Helpless | Accepted | Neutral | Calmly | Attention-grabbing | Influential |

Note: Values are scored values.

## 2.2 PAD labeling experiments

### 2.2.1 Test subject

The PAD labeling of this database was conducted by 28 students in schools and society who did not participate in the recording of the STEM-E$^2$VA database, including 15 boys and 13 girls. Since "neutral" is a kind of reference emotion, there is no emotion orientation itself, and there is no positive and negative characteristics of emotion, so the "neutral" emotion data is not used in the dimension labeling experiment. The PAD annotation experiment was divided into two parts, and a total of 36 speech data were used in Experiment 1, the main purpose of which was to ensure that the subjects could fully understand the whole experiment and grasp the judgment criteria, and 28 subjects were screened according to the experimental results. Considering the reduction of the annotation workload, we selected 8 texts with higher accuracy as Experiment 2 according to the correctness of the annotators on 16 texts in Experiment 1, in which the texts of happy, happy, angry, indifferent, sad, and sad emotions are 175, 176, 175, 176, 174, and 176, respectively.

### 2.2.2 Experimental procedure

The PAD annotation experiment platform is a small program independently developed on WeChat. Figure 7 shows the operation interface. The annotator listens to the played voice and then selects options A-G (1-7 points) based on his or her own judgment. The operation settings are as follows:

1) Three types of data are collected in the experiment. The first type is the score of the current playback statement (option); the second type is the personal response time, that is, the interval between the end of playback and the final submission; the third type is the number of clicks to repeat playback.

2) When playing the voice, the system randomly plays the voices in 6 emotions and does not prompt the emotion category of the currently played voice on the operation interface. The subjects are required to subjectively feel which emotion the current voice is and then make a score.

To allow the subjects to label efficiently, the three dimensions P, A and D are labeled separately, so that they can label in one dimension for a long time. At the same time, it is set to have a forced rest of 10 minutes after marking for 1 hour before continuing to mark.

Through experiment 1, we screened out 20 according to the correct rate of 28 annotators, and the ratio of male to female was 1∶1. Among them, the data obtained by the annotator for the first emotion is expressed as

$$e_{mi} = \begin{bmatrix} (x_{ij}^{m1}, Rt_{ij}^{m1}, Pt_{ij}^{m1}) \\ (x_{ij}^{m2}, Rt_{ij}^{m2}, Pt_{ij}^{m2}) \\ (x_{ij}^{m3}, Rt_{ij}^{m3}, Pt_{ij}^{m3}) \end{bmatrix}, \quad (1)$$

Where $\left(x_{ij}^{md}, Rt_{ij}^{md}, Pt_{ij}^{md}\right)$ represents the labeling value, reaction time and number of repetitions of the $d$ dimension on the $j$ th sample of the $i$ th emotion by the $m$ th annotator, $x$, $Rt$, and $Pt$ represent the labeled value, reaction time, and number of repetitions respectively; $m = 1, 2, \cdots, 20$, Corresponding to 20 annotators; superscript $d = 1, 2, 3$ represent pleasure, arousal and advantage respectively, $j = 1, 2, \cdots, n$, The number of samples corresponding to the current emotion, $i = 1, 2, \cdots, 6$, Corresponding to the six emotions of happiness, pleasant, anger, displeasure, sadness, and sorrow, all data labeled with person $m$ are finally expressed as:

$$E_{mi} = \{e_{m1}, e_{m2}, e_{m3}, e_{m4}, e_{m5}, e_{m6}\}。 \quad (2)$$

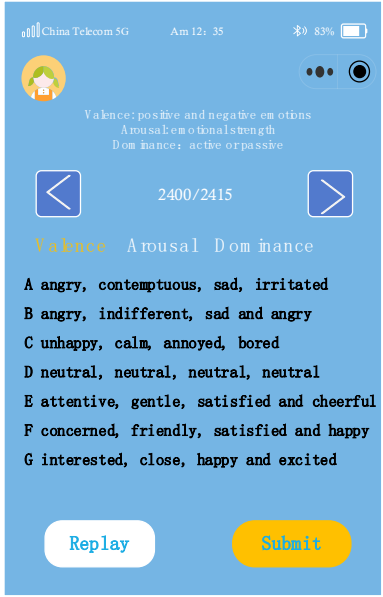

Fig.7　PAD labels WeChat mini programs

### 2.3　Data statistical analysis

Through Experiment 2, we obtained a large amount of PAD annotation data. In order to verify the validity of PAD data for subsequent research, we analyzed the discrete degree and three-dimensional spatial distribution of emotion annotation data respectively.

#### 2.3.1　Dispersion degree of annotation data

Firstly, observe the overall situation of the labeling of each emotion by 20 annotators, and describe the concentration of the labeling results by calculating the within-class standard deviation of the six emotions. The standard deviation reflects the degree of dispersion of a data set. It is the average distance of each data from the mean. The standard deviation of the same emotion coordinate distribution reflects the degree of dispersion of the emotion. The calculation formula is:

$$\sigma_i^d = \sqrt{\frac{\sum_{j=1}^{n}(x_{ij} - \overline{x_i^d})^2}{n}}。 \quad (3)$$

Where $\sigma_i^d$ is the standard deviation of the $d$ dimensional labeling result of the $i$ th emotion; $n$ is the number of samples under the $i$ th emotion; $\overline{x_i^d}$ represents the average coordinate of the $d$ th emotion in the h-dimension. Each speech sample has three-dimensional label values, so the calculated standard deviation is also three-dimensional, belonging to pleasure, arousal and dominance respectively.

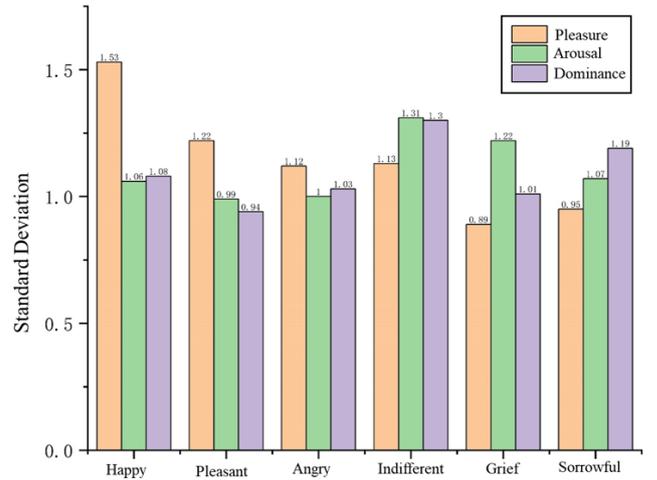

Fig.8　Standard deviation of individual emotions

The standard deviations of the six target emotions are obtained in Figure 8. The closer the standard deviation is to 0, the smaller the degree of dispersion of the emotion on the corresponding dimension; conversely, the closer the standard deviation is to 1, the greater the degree of dispersion of the emotion on the corresponding dimension. Figure 8 indicates that the standard deviation of each emotion is between 0.90 and 1.50, indicating that the data is not centrally concentrated and has a large degree of dispersion. The standard deviation of happiness is the largest, reaching 1.53. It is speculated that the individual differences of some standardizers lead to the occurrence of some outliers, resulting in a larger overall standard deviation.

### 2.3.2 Three-dimensional space distribution

Each sample of the PAD annotation data is averaged, that is, under all samples of a certain emotion, the PAD data of the 20 annotators are averaged. Due to the approximation of strong and weak emotions, it is difficult to distinguish them in their distribution in space, so we select three types of strong emotions: happiness, anger, and sadness for representation (Figure 9). It can be seen from the scatter plot that the three types of emotions can be well distinguished in the PAD three-dimensional space, and the annotated PAD data are basically distributed around the center point. In addition, we performed a 95% ellipsoid confidence interval display for the three types of emotions in three-dimensional space, and it can be seen that the three types of emotions have obvious regional concentration. At the same time, it can be seen from the semi-axis lengths of the ellipsoid in the $X, Y, Z$ (pleasure, arousal, dominance) directions of the three-dimensional coordinate space that happiness and anger are in the $X$ (pleasure) direction and sadness is in the $Y$ (arousal) direction, the ellipsoidal confidence interval becomes more elongated in that direction, which also corresponds to the standard deviation distribution in the previous section. However, the existence of outliers will have a great impact on data analysis. From the spatial distribution of experimental data, it can be seen that the annotated data obtained is concentrated in one area as a whole, and there is a clear distinction between emotions. However, due to individual differences among annotators, there are a few outliers in the spatial distribution, which is also the reason for the large degree of data dispersion in the previous section.

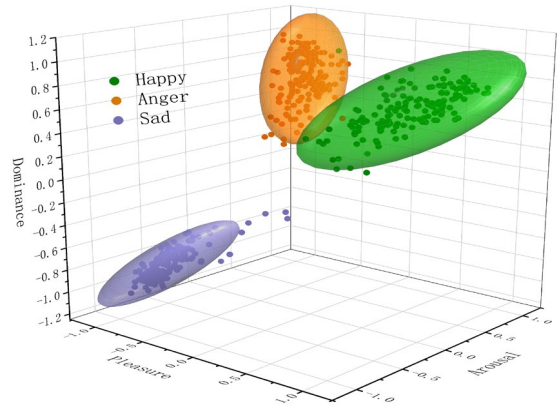

Fig.9  Spatial distribution of emotions of happy, anger, and sad

## 3 Outlier analysis and processing

The outlier phenomenon that appears in the PAD annotation experiment is due to the misunderstanding between the emotional intensity expressed by the subject and the emotional intensity felt by the annotator, especially in the emotional intensity distinction between strong and weak emotions. PAD annotation data with outliers will have a certain impact on downstream tasks, so it is necessary to filter outliers and conduct in-depth analysis of the occurrence of outliers.

### 3.1 Outlier filtering and handling

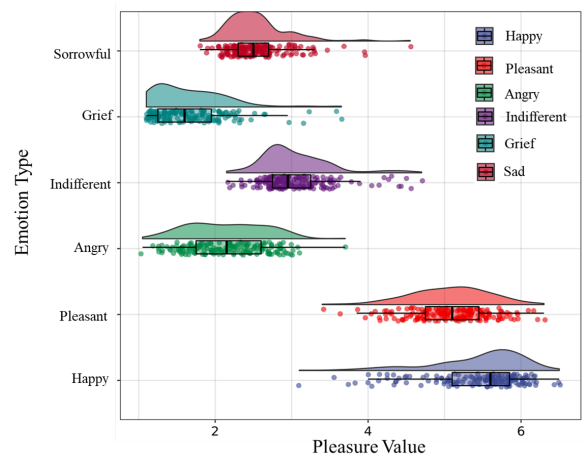

Fig.10  The average distribution of emotions

Based on the calculated average annotation value, the distribution of the six emotions in three dimensions was statistically analyzed using density estimation curves, box plots, and jitter scatter plots. The cloud and rain diagram distribution (Figure 10) can be seen that the six types of emotions all obey a skewed normal distribution. Among them, pleasant, indifferent, sad, and grief have the best concentration. The distribution of angry and happy are more scattered, but the two are obviously concentrated in a range. Therefore, the threshold $t_i^d$ of outliers is determined based on the box plot. If the threshold $t_6^1$ of sad in happy is >3.3, it is regarded as an outlier. By analogy, we can derive the thresholds of outliers in the three dimensions of the six types of emotions.

Finally, the annotation values that originally exceeded the threshold in each annotator were uniformly modified to a critical value (threshold), and the average and standard deviation of the PAD data obtained by scoring each emotion were calculated to obtain the PAD value of each emotion, as shown in Table 6. There are obvious differences in the mean values of the three strong emotions of happiness, anger, and sadness. There is also a certain gap between strong emotions and their corresponding weak emotions, and the standard deviation has also decreased overall, indicating that the PAD annotation experimental data is stable, and Reliability is better. Therefore, a dimensional emotion database can be established based on this data, and then dimensional space emotion speech recognition can be carried out.

Table. 6 PAD values for various emotions

| Emotion Category | Pleasure | Arousal | Dominance |
|---|---|---|---|
| Happy | （5.71,1.01） | （5.77,0.78） | （5.68,0.76） |
| Pleasure | （5.26,0.83） | （5.42,0.65） | （5.33,0.63） |
| Anger | （2.04,0.84） | （6.06,0.82） | （5.99,0.79） |
| Indifferent | （2.81,0.70） | （3.95,1.12） | （4.14,1.07） |
| Grief | （1.59,0.64） | （2.92,0.38） | （1.65,0.72） |
| Sad | （2.38,0.60） | （2.24,0.74） | （2.45,0.79） |

Note: The format of the values in parentheses is (mean, standard deviation)

### 3.2 SCL-90 Analysis

The Self-Rating Symptom Checklist 90 (SCL-90) is a psychosomatic screening scale proposed by Derogatis[29] and is used to assess the psychological symptom status of individuals from "healthy controls" to "disordered controls". The questionnaire consists of 90 questions covering 9 symptom dimensions (depression, hostility, anxiety, phobia, obsessive-compulsive disorder, interpersonal sensitivity, somatization, paranoia, and psychoticism). For outliers in dimension annotation values, the annotator's SCL-90 data was analyzed to explore the correlation between outliers and psychological symptoms. In order to verify whether collecting annotators' SCL-90 data is valid, we will conduct reliability and validity analyses. Reliability analysis uses the following two methods: 1) Use the Spearman-Brown formula to calculate the half-half reliability $r'_{xx}$; 2) Cronbach's coefficient $\alpha$, the formula is as follows:

$$r'_{xx} = \frac{2r_{xx}}{1+r_{xx}} \tag{4}$$

$$\alpha = \frac{n}{n-1}(1 - \frac{\sum S_i^2}{ST^2}) = \frac{n}{n-1}(1 - \frac{n}{n+2r}) \tag{5}$$

In the formula: $n$ is the half-half reliability; $r_{xx}$ is the correlation coefficient between the two halves; $n$ is the number of questions; $S_i^2$ is the witin-question variance of question $i$; $ST^2$ is the total score variance; $i$ is the sum of correlation coefficients between questions. The results are shown in Table 7. The results of the reliability study show that except for hostility and paranoia, the $\alpha$ coefficients are all above 0.8. However, the $\alpha$ of hostility and paranoia are also above 0.7 and the split-half reliability is above 0.6, indicating that the scale has good reliability.

According to psychometric theory, the correlation between each subscale score and the total scale score exceeds the correlation between each subscale, which is a test method for construct validity. We analyzed the subscale scores and the total scale score, each The correlations between the subscales are shown in Table 8. The correlation coefficient between each subscale and the total scale is between 0.640 and 0.925, and the correlation coefficient between each subscale is between 0.474 and 0.855. The correlation between the subscale and the total scale is generally greater than the correlation with the subscale. This shows that the construct validity of the scale is also relatively good, and therefore the data from the annotator SCL-90 scale obtained on this occasion can be used for further analyses.

Table. 7  Reliability analysis of SCL-90 scale

| Factor | α | $r'_{xx}$ |
|---|---|---|
| Somatization | 0.844 | 0.632 |
| Obsessive-Compulsive Disorder | 0.820 | 0.768 |
| Interpersonal sensitivity | 0.805 | 0.883 |
| Depression | 0.917 | 0.898 |
| Anxiety | 0.847 | 0.845 |
| Hostility | 0.767 | 0.819 |
| Fear | 0.809 | 0.853 |
| Paranoid | 0.729 | 0.678 |
| Psychotic | 0.852 | 0.849 |
| Overall | 0.977 | 0.966 |

### 3.3 Outlier analysis

The outlier rate of 20 annotators in each of the three dimensions of each emotion was filtered by a threshold $t_i^d$ for each emotion in each dimension. The outlier rate is equal to the number of subject annotations above the threshold divided by the total number of annotations. From this, we can derive the outlier rate for each subject in each of the three dimensions of each emotion. The state of the annotator's psychological symptoms directly affects the level of regulation of the autonomic nervous system and the hypothalamic-pituitary-adrenal axis. As a result, people with a high degree of neuroticism are more sensitive and responsive to negative stimuli and less responsive to positive stimuli. This also affects the level of emotion perception and arousal to some extent [30]. Therefore, we categorised the three dimensions of the six emotions into high and low dimensions, as shown in Table 9, to directly analyse the relationship between psychological symptoms and the high and low dimensional criteria. Within a certain range, the annotators can be allowed a certain outlier rate. As shown in Figure 11, it can be seen that the outlier rate is relatively low and stable in all categories except high pleasantness and low arousal. In the high pleasantness category, the annotators had significantly higher error rates. Four of the annotators had outlier rates of more than 30%, with a maximum of 51% (i.e., more than half of the annotated values for that annotator exceeded the threshold). Therefore, each category will be analysed in depth to explore the reasons for the outliers.

Table. 9  Emotional division details

| Dimension | Intensity | Emotion |
|---|---|---|
| Pleasure | high | happy, pleasantful |
|  | low | anger, indifference, sadness |
| Arousal | high | happiness, pleasant, anger, indifference |
|  | low | sadness |
| Dominance | high | happiness, pleasant, anger, indifference |
|  | low | sadness |

Table. 8  Spearman correlation coefficient between the aggregate table and the component table

| Factor | Somatization | Obsessive-copulsive | Interpersonal sensitivity | Depression | Anxiety | Hostility | Fear | Paranoid | Psychotic | total score |
|---|---|---|---|---|---|---|---|---|---|---|
| Somatization | 1.000 | | | | | | | | | |
| Obsessive-copulsive | 0.474 | 1.000 | | | | | | | | |
| Interpersonal sensitivity | 0.589 | 0.631 | 1.000 | | | | | | | |
| Depression | 0.643 | 0.807 | 0.785 | 1.000 | | | | | | |
| Anxiety | 0.766 | 0.575 | 0.663 | 0.768 | 1.000 | | | | | |
| Hostility | 0.659 | 0.775 | 0.834 | 0.788 | 0.612 | 1.000 | | | | |
| Fear | 0.524 | 0.436 | 0.655 | 0.652 | 0.693 | 0.607 | 1.000 | | | |
| Paranoid | 0.624 | 0.762 | 0.821 | 0.822 | 0.795 | 0.855 | 0.725 | 1.000 | | |
| Psychotic | 0.561 | 0.618 | 0.616 | 0.757 | 0.733 | 0.578 | 0.634 | 0.686 | 1.000 | |
| total score | 0.640 | 0.847 | 0.841 | 0.925 | 0.735 | 0.894 | 0.731 | 0.924 | 0.793 | 1.000 |

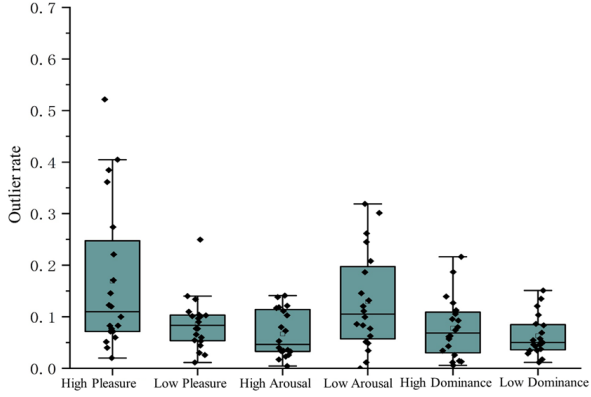

Fig.11　Outlier distribution

### 3.4 Multiple regression analysis

In order to explore the relationship between the 9 factors in psychological symptoms and 6 high and low dimensional standards, we set the 9 independent variables as somatization, obsessive-compulsive symptoms, interpersonal sensitivity, depression, anxiety, hostility, terror, paranoia, and psychosis. is $X_1 - X_9$, and the six high and low dimension standards of high and low pleasure, high and low arousal, and high and low dominance (denoted as $Y_1 - Y_6$) are the dependent variables, and multiple linear regression analysis is performed. Therefore, the preliminary model can be set as

$$Y_i = \beta_0 + \sum_{k=1}^{9} \beta_k X_k + \mu \qquad (6)$$

Where $\beta_0$ is a constant; $\beta_1, \beta_2, ..., \beta_9$ is the regression coefficient; $\mu$ is the remaining residual, which has nothing to do with the 9 independent variables. Since not all independent variables have statistically significant effects on the dependent variable, we use the stepwise regression analysis method to individually judge the significance of the 9 factors (independent variables) and 6 categories (dependent variables) of SCL-90. When the introduced independent variables become no longer significant due to the new independent variables later, they will be deleted, thereby ensuring that only significant independent variables are included in the regression equation before each new independent variable is introduced, and common variables are eliminated. Independent variables with higher linearity. Finally, we used SPSS23.0 to analyze the six categories, and the results of high pleasure are shown in Table 10.

Table. 10　Model fitting accuracy

| Model | R | $R^2$ | Adjusted $R^2$ | Error of standard estimates |
|---|---|---|---|---|
| 1 | 0.762[a] | 0.581 | 0.558 | 0.094 |
| 2 | 0.826[b] | 0.682 | 0.644 | 0.084 |
| 3 | 0.894[c] | 0.799 | 0.762 | 0.069 |

Note：a. Predictors：(constant), $X_1$, b. Predictors：(constant), $X_1$, $X_3$, c. Predictors：(constant), $X_1$, $X_3$, $X_4$。Dependent variable: High pleasure.

$R^2$ in Table 10 is the coefficient of determination, a statistic that describes the degree of fit of the regression curve to the real data points; the adjusted $R^2$ is the recalculated result of the $R^2$ coefficient of determination plus the residual degrees of freedom, indicating the fitting accuracy of the model. It can be seen from Table 10 that the adjusted $R^2$ of model 3 is 0.762, which is relatively close to 1 and has the highest fitting degree. Therefore, Model 3 is used to obtain the regression coefficient equation, as shown in Table 11. Both the unstandardized coefficient and the standardized coefficient are regression coefficients. Standardization removes dimensions, it reflects the relative importance between variables and is only used for comparison of independent variables. When interpreting the effect of independent variables on dependent variables, use unstandardized regression coefficients. In Model 3, the P<0.050 of the three independent variables $X_1$、$X_3$、$X_4$ are all significant and the tolerance of collinearity statistics is not less than 0.100, and their variance inflation factors (Variance Inflation Factor, VIF) are also less than 10.000. So, the model is available. By performing the same stepwise regression analysis on each category, the regression equation for high pleasure can be expressed as:

$$Y_1 = 0.930 + 0.398X_1 - 0.328X_3 + 0.286X_4 \qquad (7)$$

In the formula: The coefficients of somatization ($X_1$) and depression ($X_4$) are positive, which means that the more serious these symptoms are, the higher the error rate of high pleasure will be, while interpersonal sensitivity ($X_3$) shows the opposite trend. This shows that the psychological symptoms measured by the SCL-90 scale will, to a certain extent, lead to a lower level of the brain's stress response to certain emotions, which also directly affects the annotator's accuracy in the PAD annotation experiment. We used the same method to analyze the other five dependent variables, and there were no results similar to high pleasure. The reason for the outliers is unknown. This is also a direction for our subsequent analysis and research.

Table. 11 Regression equation coefficients of model3

| model3 | Non-standardized coefficient | Standardization coefficient | $t$ | $P$ | Collinear statistics | |
|---|---|---|---|---|---|---|
| | | | | | Allowance | VIF |
| (C) | 0.093 | | 3.304 | 0.004 | | |
| $X_1$ | 0.398 | 0.786 | 4.835 | 0.000 | 0.475 | 2.106 |
| $X_3$ | -0.328 | -0.790 | -4.146 | 0.001 | 0.345 | 2.895 |
| $X_4$ | 0.286 | 0.655 | 3.061 | 0.007 | 0.274 | 3.654 |

## 4 Data emotion evaluation

Dimension annotation itself has individual differences and is highly subjective. It cannot fully reflect speech quality, emotional discrimination, etc. Therefore, it is necessary to conduct objective emotion recognition evaluation on the data to verify the data of each modality of the data set. quality. As mentioned before, users can select the video data of this database according to their own needs. Therefore, this article mainly evaluates the quality of acoustic data, glottal data and articulatory kinematic data.

This article uses OpenSMLIE toolbox to extract speech features of acoustic and glottal signals - INTERSPEECH 2009 Emotional Challenge feature set (simplified to IS09 feature set) [31]. This feature set includes a total of 32 types of features, and each type of feature extracts 12 statistics. A total of 384-dimensional features and spectrogram features were obtained. In the kinematic data, the displacement and velocity characteristics of the two articulating organs of the tongue and lips during pronunciation are mainly extracted, expressed as

$$P_k = (L1_k, L2_k, L3_k, L4_k, T1_k, T2_k, T3_k) \quad (8)$$

$$V = (V_{L1}, V_{L2}, V_{L3}, V_{L4}, V_{T1}, V_{T2}, V_{T3}) \quad (9)$$

In the formula: $P_k$ represents the displacement characteristics of the left lip, right lip, upper lip, lower lip, tongue base, tongue center and tongue tip of the sensor in the $k$-axis direction; $k$ is the $X$、$Y$ and $Z$ axes respectively; $V$ is the velocity characteristics of each articulator, with a total of 28-dimensional pronunciation kinematic characteristics.

In order to test the speech emotion recognition rate of the database, we use Support Vector Machine (SVM), Deep Neural Networks (DNN) and Convolutional Neural Networks (CNN) for objective verification. Among them: SVM uses Gaussian kernel function, penalty coefficient $c = 4$, and kernel function coefficient $g = 0.25$; DNN uses 3 hidden layers with a learning rate of 0.010; CNN uses 3 layers of convolutional layers and maximum pooling layer structure, and finally goes through a full connection layer and Softmax classification layer have a learning rate of 0.001. In the entire experiment, 80% of the speech samples were used for training, and the remaining 20% of the speech samples were used for recognition. The final emotion recognition result was the confusion matrix shown in Figure 12.

The results show that the overall recognition effects of CNN, SVM and DNN are very good. The recognition rates of acoustic data reached 82.57%, 85.28% and 79.85% respectively under the three basic models; the recognition rates of glottal data reached 72.28% and 79.85 respectively. % and 65.42%; the recognition rate of pronunciation kinematic data is lower, but it also reaches 53.57%, 61.17%, and 52.28% respectively. Comparing Figure 12(b) and (c), it can be found that there is confusion between happiness and pleasure. Happiness is mostly recognized as pleasure, while pleasure is also partially recognized as happiness. This is because happiness and pleasant are a pair of strong and weak emotions, and they have many similarities in characteristics and emotions. It can be seen from Figure 12(a), (b), (c) and Figure 12(g), (h), (i) that there is obvious confusion between indifference and sadness, which shows that the vocal organs are used when expressing these two emotions. The movement and acoustic speech are similar and internally related to each other. Overall, this database can achieve a high recognition rate through simple primary features and basic recognition networks, proving that this database has high speech quality and clear distinctions between emotions, and can contribute to emotion recognition research at home and abroad. Provide good database support.

## 5 Conclusion

A multi-modal Chinese emotional speech database STEM-E$^2$VA that includes articulatory movements, speech, glottis and video are recorded in this paper. This database recruits non-professional performers to perform performance recordings which described the situation and atmosphere, trying to make up for the exaggeration and unreality of performance-based databases. The database is composed of 22 college students who expressed two types of texts: vowels and sentences.

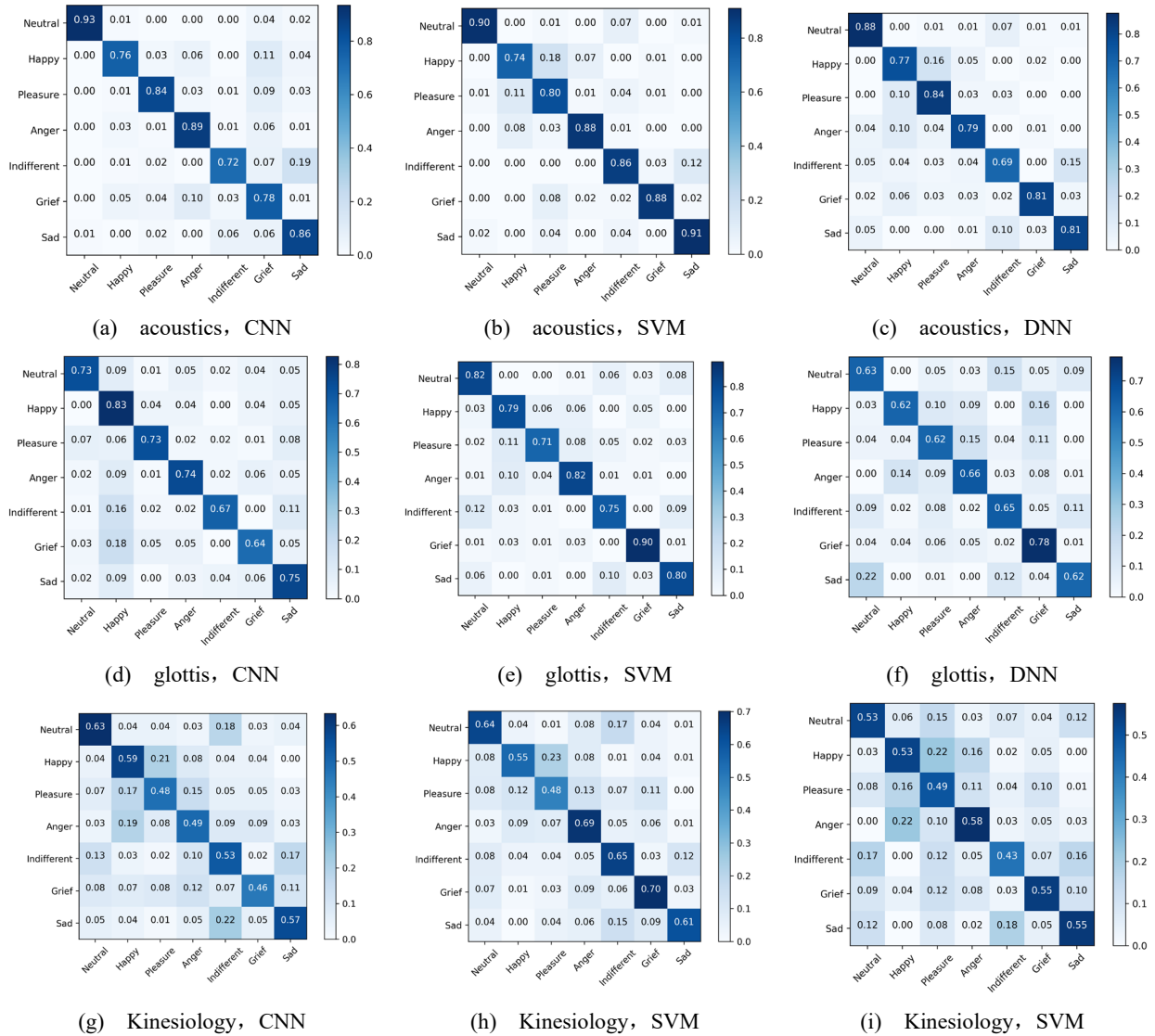

(a) acoustics，CNN  (b) acoustics，SVM  (c) acoustics，DNN

(d) glottis，CNN  (e) glottis，SVM  (f) glottis，DNN

(g) Kinesiology，CNN  (h) Kinesiology，SVM  (i) Kinesiology，SVM

Fig.12　Confusion matrices for different models

A total of 2,943 acoustic, phonetic, and laryngeal data samples and 24.5 hours of facial micro-expression data were obtained. To a certain extent, it can provide speech and articulation organs. The connections between them provide the basis for research. By dimensional labeling discrete labels, a bridge is built between discrete emotions and dimensional emotions, so that the database can also provide services for the connection between discrete labels and dimensional labels. At the same time, the discrete degree, concentration and emotional speech recognition rate of PAD data respectively verify that the database has good speech quality and emotional discrimination, and explain the inherent relationship between the outliers existing in the annotation and the annotator's psychology. In addition, this database has the following three innovations:

1) This database is the first emotional speech database at home and abroad that combines kinematic EMA data, glottal EGG data, audio data and video data.

2) The PAD scoring scale has been modified with 7 semantic points, which makes up for the inaccuracy of the 9-point semantic scale and supplements the 5-point semantic deficiencies.

3) Combine PAD labeling with the SCL-90 psychological scale to analyze the impact of the labeler's internal psychological factors on PAD labeling.

In terms of voice emotional quality, this database can achieve better standards, but there are also some shortcomings. First, due to experimental requirements, the subjects' facial micro-expression data are slightly affected by microphone occlusion and sensor sticking; second, text content and emotional scenes can be further enriched to improve the naturalness of emotional expression; finally, in addition to dimensional annotation, Emotional layer annotation can also add vowel layers, paraphonic layers, etc., thereby establishing a complete emotional speech database and providing a more complete database for speech emotion recognition at home and abroad. Finally, the face photos in the experiment have been authorized by the subjects, and their face portraits can be used.